\documentclass[12pt,epsf]{iopart}
\usepackage [dvips]{graphicx}
\usepackage{iopams}
\usepackage{epsfig}
\usepackage{verbatim}

\addtocounter{secnumdepth}{1}
\font\capfont=cmbx12 at 50 pt 
\newbox\capbox \newcount\capl \def\a{A}
\def\docappar{\medbreak\noindent\setbox\capbox\hbox{%
\capfont\a\hskip0.15em}\hangindent=\wd\capbox%
\capl=\ht\capbox\divide\capl by\baselineskip\advance\capl by1%
\hangafter=-\capl%
\hbox{\vbox to8pt{\hbox to0pt{\hss\box\capbox}\vss}}}
\def\cappar{\afterassignment\docappar\noexpand\let\a }

\begin{document}

\newcommand{\bex}{\boldsymbol{e}_x}
\newcommand{\bey}{\boldsymbol{e}_y}
\newcommand{\brr}{\mathbf{r}}
\newcommand{\bkk}{\mathbf{k}}

\newcommand{\pe}{{\cal E}}
\newcommand{\pn}{{\cal N}}
\newcommand{\px}{{\cal X}}
\newcommand{\py}{{\cal Y}}
\newcommand{\ci}{{\mathrm i}}
\newcommand{\ee}{{\mathrm e}}
\newcommand{\norm}{N}

\newcommand{\rhobar}{\overline{\rho}}
\newcommand{\rhoe }{\rho^\pe}
\newcommand{\rhon }{\rho^\pn}
\newcommand{\rhox }{\rho^\px}
\newcommand{\rhoy }{\rho^\py}
\newcommand{\rhoen}{\rho^{\pe/\pn}}
\newcommand{\rhone}{\rho^{\pn/\pe}}
\newcommand{\etabar}{\overline{\eta}}
\newcommand{\etae}{\eta^\pe}
\newcommand{\etan}{\eta^\pn}
\newcommand{\etax}{\eta^\px}
\newcommand{\etaen}{\eta^{\pe/\pn}}
\newcommand{\drhoe }{p^\pe}
\newcommand{\drhon }{p^\pn}
\newcommand{\drhox }{p^\px}
\newcommand{\drhoy }{p^\py}
\newcommand{\drhoen}{p^{\pe/\pn}}
\newcommand{\drhone}{p^{\pn/\pe}}
\newcommand{\hatrhoe}{\hat{\rho}^\pe}
\newcommand{\hatrhon}{\hat{\rho}^\pn}
\newcommand{\hatrhox}{\hat{\rho}^\px}
\newcommand{\hatrhoen}{\hat{\rho}^{\pe/\pn}}
\newcommand{\hatdrhoe}{\delta\hat{\rho}^\pe}
\newcommand{\hatdrhon}{\delta\hat{\rho}^\pn}
\newcommand{\hatdrhox}{\delta\hat{\rho}^\px}
\newcommand{\hatdrhoen}{\delta\hat{\rho}^{\pe/\pn}}
\newcommand{\Gee}{G^{\pe\pe}}
\newcommand{\Gen}{G^{\pe\pn}}
\newcommand{\Gne}{G^{\pn\pe}}
\newcommand{\Gnn}{G^{\pn\pn}}
\newcommand{\Gxy}{G^{\px\py}}
\newcommand{\gee}{g^{\pe\pe}}
\newcommand{\gen}{g^{\pe\pn}}
\newcommand{\gne}{g^{\pn\pe}}
\newcommand{\gnn}{g^{\pn\pn}}
\newcommand{\gtee}{\tilde{g}^{\pe\pe}}
\newcommand{\gten}{\tilde{g}^{\pe\pn}}
\newcommand{\gtne}{\tilde{g}^{\pn\pe}}
\newcommand{\gtnn}{\tilde{g}^{\pn\pn}}
\newcommand{\rbar}{\overline{r}}
\newcommand{\tilderhoe}{\tilde{\rho}^\pe}
\newcommand{\tilderhon}{\tilde{\rho}^\pn}
\newcommand{\tilderhox}{\tilde{\rho}^\px}
\newcommand{\tilderhoen}{\tilde{\rho}^{\pe/\pn}}
\newcommand{\ve}{v^\pe}
\newcommand{\vn}{v^\pn}
\newcommand{\ven}{v^{\pe/\pn}}

\newcommand{\bdrho}{{p}}
\newcommand{\bdrhoe}{{p^\pe}}
\newcommand{\bdrhon}{{p^\pn}}
\newcommand{\bdrhox}{{p^\px}}
\newcommand{\bdrhoen}{{p^{\pe/\pn}}}
\newcommand{\bM}{\mathbf{M}}
\newcommand{\bN}{\mathbf{N}}
\newcommand{\bMee}{\mathbf{M^{\pe\pe}}}
\newcommand{\bMne}{\mathbf{M^{\pn\pe}}}
\newcommand{\bMen}{\mathbf{M^{\pe\pn}}}
\newcommand{\bMnn}{\mathbf{M^{\pn\pn}}}
\newcommand{\bveta}{\mathbf{\eta}}
\newcommand{\bvetae}{\mathbf{\eta^\pe}}
\newcommand{\bvetan}{\mathbf{\eta^\pn}}

\newcommand{\bI}{\mathbf{I}}
\newcommand{\bA}{\mathbf{A}}
\newcommand{\bB}{\mathbf{B}}
\newcommand{\bE}{\mathbf{E}}
\newcommand{\bF}{\mathbf{F}}
\newcommand{\btF}{\mathbf{\tilde{F}}}

\newcommand{\zetag}{\zeta^\mathrm{g}}
\newcommand{\bAg}{\mathbf{A}^\mathrm{g}}
\newcommand{\bBg}{\mathbf{B}^\mathrm{g}}
\newcommand{\bEg}{\mathbf{E}^\mathrm{g}}
\newcommand{\bFg}{\mathbf{F}^\mathrm{g}}
\newcommand{\btFg}{\mathbf{\tilde{F}}^\mathrm{g}}

\newcommand{\gfdrhoen}{\hat{p}^{\pe/\pn}}
\newcommand{\gfbdrho}{{\hat{p}}}
\newcommand{\gfbdrhoe}{{\hat{p}^\pe}}
\newcommand{\gfbdrhon}{{\hat{p}^\pn}}
\newcommand{\gfbveta}{\mathbf{\hat{\eta}}}
\newcommand{\gfbvetae}{\mathbf{\hat{\eta}^\pe}}
\newcommand{\gfbvetan}{\mathbf{\hat{\eta}^\pn}}

\newcommand{\phil}{\phi^\lambda}
\newcommand{\psil}{\psi^\lambda}
\newcommand{\phim}{\phi^\mu}
\newcommand{\psim}{\psi^\mu}
\newcommand{\phitl}{\mathbf{\tilde{\phi}^\lambda}}
\newcommand{\psitl}{\mathbf{\tilde{\psi}^\lambda}}
\newcommand{\phitm}{\mathbf{\tilde{\phi}^\mu}}
\newcommand{\psitm}{\mathbf{\tilde{\psi}^\mu}}

\newcommand{\lambdat}{\mathbf{\tilde{\lambda}}}
\newcommand{\phit}{\mathbf{\tilde{\phi}}}
\newcommand{\psit}{\mathbf{\tilde{\psi}}}

\newcommand{\DQ}{{S}}
\newcommand{\cG}{{\cal G}}
\newcommand{\zet}{z_\eta}
\newcommand{\zetaet}{\zeta_\eta}
\newcommand{\epsiloneta}{\epsilon\eta}
\newcommand{\eeab}{\epsilon\eta\alpha\beta}
\newcommand{\cGp}{\cG}
\newcommand{\cGab}{\cGp_{\alpha\beta}}
\newcommand{\cGabet}{\cGp_{\alpha\beta}^\eta}
\newcommand{\Rs}{R^\mathrm{s}}
\newcommand{\Ss}{S^\mathrm{s}}
\newcommand{\Rth}{\mathcal{R}_\theta}
\newcommand{\Rsth}{R^\mathrm{s}_\theta}
\newcommand{\Ssth}{S^\mathrm{s}_\theta}
\newcommand{\cGsth}{\cG^\mathrm{s}_\theta}
\newcommand{\vc}{v_{\rm c}}
\newcommand{\vg}{v_{\rm g}}
\newcommand{\vph}{v_{\rm ph}}
\newcommand{\V}{{V}}
\newcommand{\pp}{{\prime\prime}}

\newcommand{\Rsz}{\Rs_0}
\newcommand{\DQs}{\DQ^{\mathrm{s}}}
\newcommand{\cGs}{\cG^\mathrm{s}}
\newcommand{\cGps}{\cGp^\mathrm{s}}
\newcommand{\dR}{{\delta R}}
\newcommand{\dS}{{\delta S}}

\newcommand{\la}{\langle}
\newcommand{\ra}{\rangle}
\newcommand{\beq}{\begin{equation}}
\newcommand{\eeq}{\end{equation}}
\newcommand{\bea}{\begin{eqnarray}}
\newcommand{\eea}{\end{eqnarray}}

\newcommand{\dd}{\mathrm{d}}
\newcommand{\tfrac}{\frac}
\newcommand{\cof}{\mathrm{cof}\,}

\newcommand{\gk}{\overline{k}}
\newcommand{\gvg}{\overline{v}_{\rm g}}
\newcommand{\gvph}{\overline{v}_{{\rm ph},{\rm 0}}}
\newcommand{\glambda}{\overline{\lambda}}

\title{Stripe formation instability in crossing traffic flows}

\author{{J. Cividini and H.J.~Hilhorst}\\[5mm]
{\small Laboratoire de Physique Th\'eorique, b\^atiment 210}\\
{\small Universit\'e Paris-Sud and CNRS,
91405 Orsay Cedex, France}\\}

\date{\today}

\begin{abstract}
At the intersection of two unidirectional
traffic flows a stripe formation instability is known to occur. 
In this paper we consider coupled time evolution equations 
for the densities of the two flows in their intersection area.
We show analytically how the instability arises from
the randomness of the traffic entering the area.
The Green function of the linearized equations is shown to form a Gaussian wave packet whose oscillations correspond to the stripes. Explicit formulas are obtained for various characteristic quantities in terms of the traffic density and comparison is made with the much simpler calculation on a torus and with numerical solution of the evolution equations.
\end{abstract}

\maketitle


\section{Introduction} 
\label{section:introduction}

\begin{figure}
\begin{center}
\scalebox{.50}
{\includegraphics{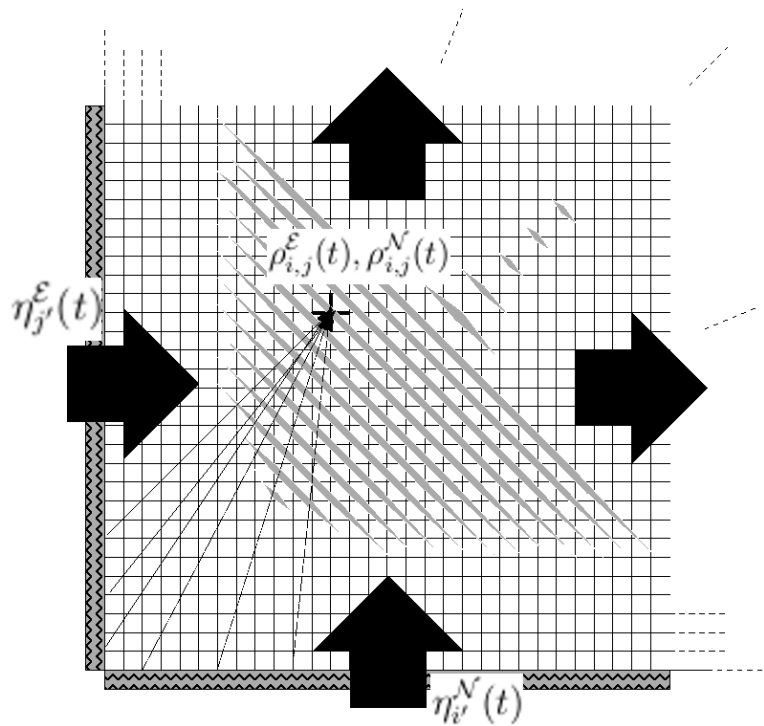}} 
\end{center}
\caption{\small Schematic representation of the square region where the crossing flows interact. The boundary
  noise  $\etae_{j'}(t)$ and $\etan_{i'}(t)$ is applied at the west and
  south boundaries of an $M \times M$ square grid. The perturbations of
  the density fields $\rhoe_{i,j}(t)$ and $\rhon_{i,j}(t)$ propagate eastward and northward according to
  Eqs.\,(\ref{eq:nl}), and exit the system at the east and north
  boundaries. The stripe instability is shown.}  
\label{fig:scheme}
\end{figure}

In traffic dynamics,
crossing flows, whether of pedestrians or of vehicles,
have attracted a certain amount of attention in recent years.
The crossing of two single lanes was studied, for example, 
in Refs.\,\cite{nagel_s1992,fouladvand_s_s2004b,foulaadvand_n2007,%
du2010,appert-rolland_c_h2011c}.
Here we will turn our interest towards wider lanes, that have been the object of experimental studies on pedestrians \cite{naka1977, ando_o_o1988, hoogendoorn_d2003, plaue2011, bamberger2014} and for which realistic models have been designed \cite{hoogendoorn_b2003, yamamoto_o2011}. Monte Carlo studies of simpler cellular automaton models of such intersecting flows were carried out by several groups \cite{biham_m_l1992, tadaki1996, watanabe2003a, watanabe2003b, fukui_i2010, xiaomei2011, ding_j_w2011, ding2011, ding2012, sui2012, muramatsu_n2000} including ourselves \cite{cividini_a_h2013,cividini_h_a2013,cividini_a2013,hilhorst_c_a2014}.
In most of the studies cited a vehicle or a pedestrian, as the case may be, is
represented by a hard core particle on a lattice site.
It is known from simulations \cite{biham_m_l1992,hoogendoorn_b2003, yamamoto_o2011,
ding2012,moussaid2012}
and from
experiments \cite{naka1977, ando_o_o1988, dzubiella_h_l2002, hoogendoorn_d2003, plaue2011, zhang_s2014}
that when two unidirectional flows cross,
whether perpendicularly or at an angle,
there arises a stripe formation instability. 
In the case of perpendicular flows,
in the square region where the flows intersect
the two kinds of particles 
show a pattern of alternating stripes
approximately or exactly perpendicular to the $(1,1)$ direction, as shown in Fig.\,\ref{fig:scheme}.
It is the purpose of this work better to understand
this stripe formation instability in perpendicularly crossing flows.
\vspace{2mm}

The analytic approach to this problem, and in fact to almost any question
concerning crossing particle flows, is very hard: these are strongly
interacting many-particle systems. 
As a simplification we introduce two continuous fields $\rhoe_{i,j}(t)$ and
$\rhon_{i,j}(t)$  ($\pe$ for eastbound and $\pn$ for northbound)
representing the densities of the two species
at times $t=0,1,2,\ldots$ on the lattice sites $i,j=1,2,\ldots,M$
that represent the intersection square.  
Then, largely independently of the precise microscopic rules of motion
of the particles, one postulates the time evolution equations
\bea
\label{eq:nl}
\rhoe_{i,j}(t+1) &=& (1-\rhon_{i,j}(t)) \rhoe_{i-1,j}(t) +
\rhon_{i+1,j}(t) \rhoe_{i,j}(t), \nonumber\\[2mm] 
\rhon_{i,j}(t+1) &=& (1-\rhoe_{i,j}(t)) \rhon_{i,j-1}(t) +
\rhoe_{i,j+1}(t) \rhon_{i,j}(t), 
\eea
whose boundary conditions we will discuss shortly.
These equations are believed \cite{cividini_a_h2013,cividini_h_a2013}
to be representative of the class of
unidirectional deterministic particle dynamics at sufficiently low density, irrespective of the
exact details of the evolution at the particle level.
In the absence of the nonlinear terms all particles would simply cross
the square at unit velocity without any impediment.
The terms in (\ref{eq:nl}) that are quadratic in the densities
express that an $\pe$ particle
that tries to hop forward will be blocked if there is a $\pn$ particle
on its target site, and the other way around. Blockings between
same-type particles are expected
to correspond to higher order effects in the density~\cite{cividini_h_a2013}
and are neglected in this description.

Eqs.\,(\ref{eq:nl}) have to be supplied with initial and
boundary conditions.
Following the example of the BML model \cite{biham_m_l1992} several authors
have studied crossing flows with periodic boundary conditions. 
If one adopts periodic boundary conditions (PBC) equations (\ref{eq:nl}) are translationally invariant in
both the $i$ and the $j$ direction and therefore
allow for a uniform stationary state in which
$\rhoe_{i,j}(t)=\rhon_{i,j}(t)=\rho$ for all $i,j$, with a value of $\rho$ determined by the initial condition.\footnote{Under periodic boundary conditions the total mass of $\pe$
particles ($\pn$ particles) in each row (column) is conserved, 
so that there are obviously many other stationary states.}
However,
a linear stability analysis 
shows that this stationary
state is unstable to random perturbations of the initial condition (Ref.\,\cite{cividini_h_a2013}, section 4). 
One of the few analytic results in this field  
is an expression for the wavelength 
and the growth rate of the most unstable mode
as a function of the density.

The true problem of crossing flows, however, has open boundary
conditions (OBC) and is driven by a
random inflow of particles at its western and southern boundaries.
Whereas the calculation with periodic boundary conditions 
does make the observed instability plausible,
the question nevertheless remains whether random boundary conditions,
rather than random initial conditions, lead to the same instability.
In this paper we address this problem. We do so
again by linearizing Eqs.\,(\ref{eq:nl}),
but now under random boundary conditions at the two entrance
boundaries. 
Specifically, we will use Eqs.\,(\ref{eq:nl}) for $i,j=1,2,\ldots,M$
with the stipulation that
\bea\label{eq:nlbc}
\rhoe_{0,j}(t) = \rho+\etae_j (t), \qquad j=1,\ldots,M, \nonumber\\[2mm]
\rhon_{i,0}(t) = \rho+\etan_i (t), \qquad i=1,\ldots,M,
\eea
in which $\etae_j (t)$ and $\etan_i (t)$ are noise terms of zero mean
that express that the particles enter randomly; these terms
may be associated with the `entrance sites' in Fig.\,\ref{fig:scheme}.
On the exit boundaries we make the most convenient choice
$\rhoe_{i,M+1}(t)=\rhon_{M+1,j}(t)=\rho$ for all $i,j=1,\ldots,M$, keeping in mind that this choice has very little influence on the physical properties of the system.

After an analysis of considerable complexity we find that
the random boundary conditions (\ref{eq:nlbc}), 
too, lead to a stripe formation instability. 
We compare the expression for its $\rho$ dependent growth rate and
maximally unstable wavelength with those found under periodic boundary conditions in Refs.\,\cite{cividini_a_h2013,cividini_h_a2013}
and find -- which was far from obvious {\it a priori} --
that they are identical. 
The stripe formation instability therefore appears to be an intrinsic property of the equations.

\begin{figure}
\begin{center}
\scalebox{.50}
{\includegraphics{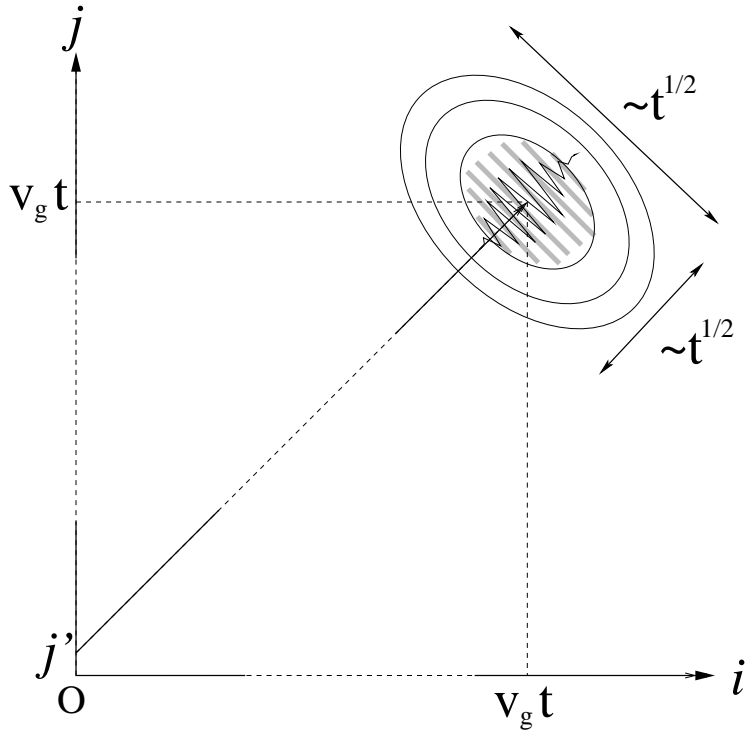}} 
\end{center}
\caption{\small Typical shape of the Green functions. Here the $\pe$ density has been perturbed on the boundary site $(1,j')$ at time $t'=0$. After a long enough time $t$ the Green function consists in a Gaussian wavepacket centered at $(i,j) = (\vg t, \vg t)$, where the velocity $\vg$ is determined in Eq. (\ref{eq:rvg}). The represented wave packet is wider in the direction perpendicular to the diagonal in accordance with equations (\ref{eq:sigmaparpeak}) and (\ref{eq:sigmaperppeak}).}  
\label{fig:wavepacket}
\end{figure}

Our work furnishes, moreover, a new look onto the problem.
We find that an instantaneous and localized perturbation
applied at a boundary site 
and superposed on a uniform background of density $\rho$
propagates inward along a diagonal at a group velocity $\gvg$
that we are able to determine as a function of the 
background density $\rho$.
This propagating pulse widens diffusively, hence acquiring a Gaussian envelope. We are able to calculate its widths along and perpendicularly to the direction of propagation. 
In addition, the propagating pulse shows oscillations that we fully characterize analytically, thereby demonstrating that stripe formation indeed occurs.
The structure and dimensions of the pulse are shown schematically in figure \ref{fig:wavepacket}.

\vspace{2mm}

This paper is organized as follows.
In section\;\ref{section:exact} we 
linearize Eqs.\,(\ref{eq:nl}) and obtain a system
governed by a $2M^2\times 2M^2$ time
evolution matrix. Green functions are defined for each of the four
$M^2 \times M^2$ subblocks.
The equations are solved in terms of generating functions
in subsection\;\ref{subsection:formal}.
This solution is partially formal and involves $M^2\times M^2$ matrices 
$\bE$ and $\bF$. 
These matrices are made explicit
in subsection\;\ref{subsection:diag}, where we also carry out the
required diagonalization of $\bF$. 
In subsection\;\ref{subsection:exprgreen} we combine the preceding results and obtain fully explicit exact expressions for the four
Green functions for finite $M$, which take the form of an inverse Fourier-Laplace
transform.
In section\;\ref{section:inversion} we perform
an asymptotic expansion valid for large times
and distances and calculate the properties of the propagating wave packet. 
The expansion starts with finding, in
subsection\;\ref{subsection:poles}, the poles
of the Green function in the plane of the variable $z$ conjugate to time.
In subsection\;\ref{subsection:dompole} we select the pole expected to
dominate in the large time limit.
The asymptotic analysis then becomes a
saddle-point calculation in the planes of the Fourier variables. The general structure of this calculation is discussed in 
subsection\;\ref{subsection:general}.
The wave packet is studied explicitly along the diagonal in
subsection\;\ref{subsection:ondiag} and in the vicinity of the diagonal
 in subsection\;\ref{subsection:offdiag}.
Section\;\ref{section:ccl}
summarizes the results and concludes the paper. 


\section{Linearized equations}
\label{section:exact}

In this section we will study the linearized version of the time evolution equations (\ref{eq:nl}).
We write $\rhox_{i,j}(t) = \rho+\drhox_{i,j}(t)$ for all
$1\leq i,j \leq M$, where $\px=\pe,\pn$,
the $\drhox_{i,j}(t)$ are small, and $\rho$
is the average of the entrance site densities defined in Eq.\,(\ref{eq:nlbc}). 
The linearization of Eqs.\,(\ref{eq:nl}) reads 
\bea
\label{eq:l}
\drhoe_{i,j}(t+1) &=& (1-\rho) \drhoe_{i-1,j}(t) + \rho
\drhoe_{i,j}(t) - \rho \drhon_{i,j}(t) + \rho \drhon_{i+1,j}(t) ,
\nonumber\\[2mm] 
\drhon_{i,j}(t+1) &=& (1-\rho) \drhon_{i,j-1}(t) + \rho
\drhon_{i,j}(t) - \rho \drhoe_{i,j}(t) + \rho \drhoe_{i,j+1}(t)
\eea 
for $1\leq i,j \leq M$ and $t=0,1,2,\ldots$ for all $M \geq 1$.
The entrance boundary conditions (\ref{eq:nlbc}) become
\bea
\label{eq:bcin}
\drhoe_{0,j}(t) = \etae_j (t), \qquad j=1,\ldots,M,\nonumber\\[2mm]
\drhon_{i,0}(t) = \etan_i (t), \qquad i=1,\ldots,M.
\eea
and the exit boundary conditions are 
\beq
\label{eq:bcout}
\drhoe_{i,M+1}(t)=\drhon_{M+1,j}(t)=0 \qquad i,j=1,\ldots,M.
\eeq
We will take the system at the initial time $t=0$
in a state of uniform density $\rho$, that is, 
\bea
\label{eq:ic}
\drhox_{i,j}(0) = 0, \qquad \px=\pe,\pn.
\eea
Eqs.\,(\ref{eq:l}), (\ref{eq:bcout}), and (\ref{eq:ic}) are homogeneous in the $\drhox_{i,j}(t)$ so that the whole system (\ref{eq:l})-(\ref{eq:ic}) would have only the zero solution if the entrance noises $\etae_j(t)$ and $\etan_i(t)$ both vanished.
The problem (\ref{eq:l})-(\ref{eq:ic}) is linear, and its solution may therefore be written
as a convolution of the time dependent boundary noise
with an appropriate Green function.
Given a unit
perturbation applied on a boundary site $(i',0)$ or $(0,j')$
at some time $t'$
to one of the two entering fluxes, the Green function tells us
the effect
on the densities at arbitrary later times $t>t'$ at arbitrary
lattice sites $(i,j)$.

\subsection{Solution in terms of generating functions}
\label{subsection:formal}

Let $\bdrhoe(t)$ stand for the $M^2$-component vector
containing all values of the
fields $\drhoe_{i,j}(t)$,  and similarly  $\bdrhon(t)$ for the vector
of the $\drhon_{i,j}(t)$. 
Equations (\ref{eq:l})-(\ref{eq:bcout}) may be written vectorially 
with the aid of two $M \times M$ matrices $\bA$ and $\bB$ defined by
\beq
\label{eq:defA}
\bA_{i;i'} \equiv (1-\rho) \delta_{i;i'+1} + \rho \delta_{i;i'},
\eeq
\beq
\label{eq:defB}
\bB_{i;i'} \equiv - \rho \delta_{i;i'} + \rho \delta_{i;i'-1},
\eeq
where $\delta_{i;i'} = 1$ if $i=i'$ and $0$ otherwise.
Letting $\bI$ stand for the $M\times M$ identity matrix
we now define four $M^2 \times M^2$ matrices that act
on the tensor product space between columns $i$ and rows $j$,
\bea
\label{eq:defmat}
\bMee &\equiv& \bA \otimes \bI, \nonumber\\
\bMen &\equiv& \bB \otimes \bI, \nonumber\\
\bMne &\equiv& \bI \otimes \bB, \\
\bMnn &\equiv& \bI \otimes \bA, \nonumber
\eea
that is, componentwise,
$ [\bMee]_{i,j;i',j'} = \bA_{i;i'}\bI_{j;j'}$, {\it etc.}
Upon setting $p(t)\equiv{\bdrhoe(t) \choose \bdrhon(t)}$ and 
$\eta(t)\equiv{\bvetae(t) \choose \bvetan(t)}$
we may cast the system (\ref{eq:l})-(\ref{eq:bcout}) of linearized
equations in the form 
\beq
\label{eq:compact}
\bdrho(t+1) = \bM \bdrho(t) +(1-\rho)\bveta(t),
\eeq
in which
\beq
\label{eq:defM}
\bM \equiv \left( \begin{array}{cc} \bMee & \bMen \\ \bMne &
    \bMnn \end{array} \right),
\eeq
and with the noise vectors defined as
$(\bvetae(t))_{ij} =
\delta_{i;1} \etae_j(t)$ and $(\bvetan(t))_{ij} = \etan_i(t) \delta_{j;1}$. 

This linear equation may be solved by generating
function methods. We define the generating function 
(or Laplace transform) of any $X_t$ by
$\hat{X}(z) \equiv \sum_{t=0}^\infty z^tX_t$, where $z$ is a complex
number within the radius of convergence of the sum. 
This transformation is inverted by integrating in the complex
plane $X_t = (2 \pi \ci)^{-1} \oint_{\Gamma_0} \dd z\, z^{-t-1}
\hat{X}(z)$, where $\Gamma_0$ runs counterclockwise around the
origin. Applying this transformation to Eq.\,(\ref{eq:compact}) with
initial condition\,(\ref{eq:ic}) gives 
\beq
\label{eq:compactgf}
z^{-1} \gfbdrho = \bM \gfbdrho +(1-\rho)\gfbveta,
\eeq
where we omitted the argument $z$ of $\gfbdrho$ and $\gfbveta$.
After a little algebra one obtains
\bea 
\label{eq:pgf}
\gfbdrhoe &=& (1-\rho) (\bI \otimes \bI-\bF \otimes \bF)^{-1}  \bigg[ (\bE \otimes \bI) z
\gfbvetae + (\bF \otimes \bE) z \gfbvetan \bigg] ,\nonumber\\ 
\gfbdrhon &=& (1-\rho) (\bI \otimes \bI-\bF \otimes \bF)^{-1} \bigg[ (\bI \otimes \bE) z
\gfbvetan + (\bE \otimes \bF) z \gfbvetae \bigg]. 
\eea
where 
\beq
\label{eq:defEF}
\bE(z) \equiv (\bI-z \bA)^{-1}, \qquad \bF(z) \equiv (\bI-z \bA)^{-1} z \bB,
\eeq
and the various inverse matrices exist for almost all values of $z$.

We define the Green functions $\Gxy$ by the convolutions
\bea
\label{eq:defgreen}
\drhoe_{i,j}(t) & = & \sum_{t'=0}^{t-1} \bigg[
\sum_{j'=1}^{M} \Gee_{i,j;j'}(t-t') \etae_{j'}(t') +
\sum_{i'=1}^{M} \Gen_{i,j;i'}(t-t') \etan_{i'}(t')   \bigg],
\nonumber \\ 
\drhon_{i,j}(t) & = & \sum_{t'=0}^{t-1} \bigg[
\sum_{i'=1}^{M} \Gnn_{i,j;i'}(t-t') \etan_{i'}(t') +
\sum_{j'=1}^{M} \Gne_{i,j;j'}(t-t') \etae_{j'}(t')   \bigg]. 
\eea
The generating function may then be inverted to give
the following expressions for the Green functions in terms of the
matrices $\bE$ and $\bF$,
\bea
\label{eq:exprgreen}
\Gee_{i,j;j'}(t-t') &=& \frac{1-\rho}{2 \pi \ci} \oint_{\Gamma_0}
\frac{\dd z}{z^{t-t'}} \big[ (\bI \otimes \bI - \bF \otimes \bF)^{-1} (\bE \otimes \bI)
\big]_{i,j;1,j'}\,,\nonumber \\[2mm] 
\Gen_{i,j;i'}(t-t') &=& \frac{1-\rho}{2 \pi \ci} \oint_{\Gamma_0}
\frac{\dd z}{z^{t-t'}} \big[ (\bI \otimes \bI - \bF \otimes \bF)^{-1} (\bF \otimes \bE)
\big]_{i,j;i',1}\,.
\eea
Symmetric formulas for $\Gnn_{i,j;i'}(t-t')$ and
$\Gne_{i,j;j'}(t-t')$ are obtained
by inversion of the column and row indices. 
With these expressions we have succeeded in disentangling the four
$M^2 \times M^2$ blocks in equation (\ref{eq:compactgf}).
They remain formal within each block until we are able to explicitize the
integrands in Eqs.\,(\ref{eq:exprgreen}). This is our next task.

In order to evaluate $(\bI \otimes \bI -\bF \otimes \bF)^{-1}$ we need
to diagonalize $\bF$. 
This will be done in detail in subsection~\ref{subsection:diag},
where we show that $\bF$ has full biorthonormal
sets of right and left eigenvectors,
$\{\phil\}$ and $\{\psil\}$, respectively, associated with a set 
of eigenvalues $\{\lambda\}$.
We may therefore write $\bF = \sum_{\lambda} \psil
\lambda \phil$ where $\sum_{\lambda}$ ranges over the whole
spectrum of $\bF$. The eigenvectors satisfy $\bF \psil = \lambda \psil$ and
$\phil \bF = \lambda \phil$, as well as $\phil \cdot \psim \equiv \sum_{i=1}^M \phil_i \psim_i= \delta_{\lambda,\mu}$.  Using the diagonal form of
$\bF$ we finally get 
\beq 
\label{eq:evFEI}
\big[ (\bI \otimes \bI- \bF \otimes \bF)^{-1} (\bE \otimes \bI) \big]_{i,j;i',j'} =
\sum_{\lambda,\mu} \frac{\psil_i \psim_j
  \phim_{j'}}{1-\lambda \mu} \sum_{i'' = 1}^M \phil_{i''} \bE_{i'';i'}, 
\eeq
\beq
\label{eq:evFFE}
\big[ (\bI \otimes \bI- \bF \otimes \bF)^{-1} (\bF \otimes \bE) \big]_{i,j;i',j'} =
\sum_{\lambda,\mu} \lambda \frac{\psil_i \psim_j
  \phil_{i'}}{1-\lambda \mu} \sum_{j'' = 1}^M \phim_{j''} \bE_{j'';j'}.
\eeq
The important achievement here is that with 
Eqs.\,(\ref{eq:evFEI}) and (\ref{eq:evFFE})
we have come as near as is
possible to decoupling the motion in the two orthogonal directions:
the right hand members of each of these equations would 
factorize into an $i$ and a $j$ dependent part
if it were not for the factor $(1-\lambda\mu)^{-1}$.
This factor is a very succinct
representation in reciprocal space of the interaction between the two flows.

\subsection{Diagonalizing\, $\bF$}
\label{subsection:diag}

In order to prepare for diagonalizing $\bF$ we will first find the explicit
expressions of its matrix elements $\bF_{i;i'}$.
From Eq.\;(\ref{eq:defA}) it follows that $[\bA^n]_{i;i'} = \sum_{k=0}^n
{n \choose k}\rho^{n-k} (1-\rho)^k \delta_{i;i'+k}$ for $i,i' = 1, \ldots, M$.
We define $\zeta
\equiv \frac{(1-\rho)z}{1-\rho z}$, which has the inverse
$z = \frac{\zeta}{(1-\rho) + \rho \zeta}$. For the matrix $\bE$ we
get 
\bea
\label{eq:explE}
\bE_{i;i'} &=& \sum_{p=0}^\infty z^p [ \bA^p ]_{i;i'} \nonumber \\
&=& \Theta(i \geq i') \frac{1}{1-z \rho} \zeta^{i-i'}
\eea
with $\Theta(a) = 1$ if assertion $a$ is true and $0$
otherwise. Eq.\,(\ref{eq:explE}) is valid when the sums converge, 
{\it i.e.} for $|z| < \rho^{-1}$.
From equations (\ref{eq:defEF}) and (\ref{eq:explE}) we find
\bea
\label{eq:explF}
\bF_{i;i'} &=& [\bE z \bB]_{i;i'} \nonumber \\
&=& ( \zeta \Theta(2 \leq i' \leq i+1) - \Theta(1 \leq i' \leq i) )
\frac{z \rho}{1- z \rho} \zeta^{i-i'},
\eea
which is the desired explicit expression.

We write $\bF_{i;i'} = \frac{z \rho}{1-z \rho} \zeta^{i-i'}
\btF_{i;i'}$, where \[ \btF = \left( \begin{array}{cccc} -1 & \zeta &
    0 & \ldots \\ -1 & -1+\zeta & \zeta & \ddots \\ -1 & -1+ \zeta &
    -1+\zeta & \ddots \\ \vdots & \vdots & \vdots & \ddots \end{array}
\right) .\] The right and left eigenvectors and the eigenvalues of $\btF$ will be denoted by
$\psitl$, $\phitl$, and $\lambdat$, respectively. The eigenproperties of $\bF$ follow from those of $\btF$ by
$\psil_k = \zeta^k \psitl_k$, $\phil_k = \zeta^{-k}
\phitl_k$, and $\lambda =\frac{z \rho}{1-z \rho} \lambdat$. 

The equation for the right
eigenvector $\btF \psitl = \lambdat \psitl$ reads in
components
\beq
\label{eq:revH1}
\left\{ 
\begin{array}{l l l}
-\psitl_1 + \zeta \psitl_{2} =
\lambdat \psitl_1\,, \\[2mm] 
-\psitl_1 + (\zeta-1) \sum_{i=2}^{k} \psitl_i + \zeta \psitl_{k+1} =
\lambdat \psitl_k\,, \qquad k=2,\ldots,M-1, \\[2mm] 
-\psitl_1 + (\zeta-1) \sum_{i=2}^{M} \psitl_i =
\lambdat \psitl_M. \\
\end{array}\right.
\eeq
Subtracting the equation for $k$ from the one for $k+1$ 
for $k = 2,\ldots,M-1$ and introducing convenient boundary conditions gives, equivalently, 
\beq
\label{eq:revH2}
\left\{ 
\begin{array}{l l}
-\psitl_k + \zeta \psitl_{k+1} = \lambdat (\psitl_k-\psitl_{k-1}),
\qquad k=1,2,\ldots,M, \\[2mm] 
\psitl_0 = \psitl_{M+1} = 0. \\
\end{array}\right.
\eeq
The first equation is a linear second-order recurrence relation that
can be solved by an arbitrary linear combination of two fixed
geometric sequences. The terminal conditions provided by the second
equation fix the coefficients of this combination. Defining\footnote{Note that $\ci \equiv \sqrt{-1}$ and $i$ is the first coordinate of a lattice site $(i,j)$.} 
\beq \label{eq:defaq}
a_q \equiv \cos q + \ci (\zeta^{-1} - \cos^2q)^{1/2}
\eeq
we can write the
$M$ right eigenvectors of $\btF$ as  
\beq
\label{eq:exprpsi}
\psitl_k = \ci a_q^k (e^{\ci k q}-e^{-\ci k q}),
\qquad k=1,2,\ldots,M,
\eeq
corresponding to the eigenvalue 
\beq
\label{eq:exprlambda}
\lambdat = \zeta a_q^2 = 2 \zeta \cos q\, a_q + 1
\eeq
for $q = \frac{\pi \kappa}{M+1}$, $\kappa = 1,\ldots, M$.
Similar reasoning leads to the expression for the left eigenvectors 
\beq
\label{eq:exprphi}
\phitl_k = \frac{-\ci}{\norm_q a_q^k} \bigg[ \big( 1- a_q^{-1}
e^{- \ci q}\big) e^{-\ci k q}  - \big( 1- a_q^{-1} e^{\ci q}\big)
e^{\ci k q}\bigg], 
\eeq
where $\norm_q$ is a
normalization constant. 

We now return to the matrix $\bF$. From $\phil \cdot \psil =1$ we deduce $\norm_q = (M+1) \frac{2 \ci (\zeta^{-1} - \cos^2 q)^{1/2}}{a_q}$. As a useful intermediate result we also get 
\beq
\label{eq:explphiE}
\sum_{i'' = 1}^M \phil_{i''} \bE_{i'';i'} = \frac{2 \zeta^{-i'} \big[ a_q^{-(M+1)} \sin((M+1)q) - a_q^{-i'} \sin(i' q) \big]}{(1-z \rho) \norm_q} .
\eeq
The diagonalization of $\bF$ is now complete and the explicit results of this section should be substituted in (\ref{eq:evFEI}) and (\ref{eq:evFFE}).

\subsection{Expressions of the Green functions}
\label{subsection:exprgreen}

We are now able to bring all the pieces together to get an explicit expression for the Green functions. We define $c(\rho) \equiv \sqrt{\frac{1-\rho}{\rho}}$. Combining (\ref{eq:exprgreen}) with either (\ref{eq:evFEI}) or (\ref{eq:evFFE}) and the explicit expression of the matrix $\bE$ (\ref{eq:explE}) as well as the eigenvalues and eigenvectors of $\bF$ given by (\ref{eq:exprpsi}), (\ref{eq:exprlambda}), and (\ref{eq:exprphi}), we finally get
\bea
\label{eq:explGaM}
 \Gee_{i,j;j'}(t) &=& \frac{4}{(M+1)^2} \sum_{q,p} \frac{1}{2 \pi \ci} \oint_{\Gamma_0} \frac{\dd z}{z^{t+1}} \,g^{(M)}_{i,j;j'}(z;q,p), \\
\label{eq:explGbM}
 \Gen_{i,j;i'}(t) &=& \frac{4}{(M+1)^2} \sum_{q,p} \frac{1}{2 \pi \ci} \oint_{\Gamma_0} \frac{\dd z}{z^{t+1}} \bigg( \frac{\zeta a_{p}}{c} \bigg)^2\,g^{(M)}_{j,i;i'}(z;q,p),
\eea
where $\sum_{q,p}$ is understood as $\sum_{\kappa=1}^M \sum_{\iota=1}^M$ with $q \equiv \frac{\pi \kappa}{M+1}$ and $p \equiv \frac{\pi \iota}{M+1}$. The integrand reads
\bea
\label{eq:explgM}
g^{(M)}_{i,j;j'}(z;q,p) &\equiv&  \frac{[\sin((j'+1)p) - a_p \sin j'p][\sin q- a_q^{-M} \sin((M+1)q)]}{(\zeta^{-1} -\cos^2 q)^{1/2} (\zeta^{-1}-\cos^2 p)^{1/2}} \nonumber\\[2mm] 
&&\times \frac{(\zeta a_q)^i (\zeta a_p)^{j-j'} \sin(iq) \sin(jp) }{1-c^{-4} \zeta^4 a_q^2 a_p^2 } \,,
\eea
where we recall that $\zeta=(1-\rho)z/(1-\rho z)$
and $a_q = \cos q + \ci (\zeta^{-1} - \cos^2q)^{1/2}$.
One may check that (\ref{eq:explGaM}) and (\ref{eq:explGbM})
are real by noticing
that the symmetry operation $(\kappa,\iota) \mapsto (M+1-\kappa,M+1-\iota)$ converts
the contour integrals into their complex conjugates. 

Eq.\,(\ref{eq:explgM}) gives the exact Fourier-Laplace transforms, up to known factors, of
the Green functions 
and Eqs.\,(\ref{eq:explGaM})-(\ref{eq:explGbM}) 
are the standard inversion formulas.

\section{Inversion of the Fourier-Laplace transform}
\label{section:inversion}

The Fourier-Laplace
inversion represented by Eqs.\,(\ref{eq:explGaM})-(\ref{eq:explGbM})
can be carried out in an exact closed form only
asymptotically in the limit of large times $t$.
Since expressions~(\ref{eq:explGaM}) and~(\ref{eq:explGbM})
for $\Gee$ and $\Gen$ differ only by time-independent factors which are negligible in the $t \rightarrow \infty$ limit, we focus on $\Gee$.

We start by taking the $M \rightarrow\infty$ limit of equation (\ref{eq:explGaM}). In this limit we have $\frac{1}{M+1} \sum_q \rightarrow \frac{1}{\pi} \int_{0}^\pi dq$.
We may therefore write the Green function as
\beq
\label{eq:explGa}
\Gee_{i,j;j'}(t) = 4 \int_{0}^{\pi} \frac{\dd q}{\pi}
\int_{0}^{\pi} \frac{\dd p}{\pi} \frac{1}{2 \pi \ci}
\oint_{\Gamma_0} \frac{\dd z}{z^{t+1}} \,g_{i,j;j'}(z;q,p), 
\eeq
where $g_{i,j;j'}(z;q,p) \equiv g^{(\infty)}_{i,j;j'}(z;q,p)$ is obtained from (\ref{eq:explgM}) by removing the $M$ dependent term $- a_q^{-M} \sin((M+1)q)$.

In this section we study the large time limit, in an appropriate scaling regime, of (\ref{eq:explGa}).
We let $i, j$, and $t$ become large with
$j'$ remaining finite, \textit{i.e.} we study the
propagation of a perturbation far from the boundary where it was
created. More explicitly, we anticipate that
an instantaneous pointlike perturbation imposed at one of the
boundaries will travel 
in the $(1,1)$ direction at some yet unknown speed
while spreading diffusively.
We therefore scale $i$ and $j$ as 
\beq
\label{eq:ijscaling}
i=vt+u\sqrt{t}, \qquad j=vt-u\sqrt{t},
\eeq
where $u$ and $v$ are constants. 

It will be profitable for the developments to come
to transform the pair of variables
$(q,p)$ successively to another pair $(Q,P)$ and a third pair $(R,S)$
defined by
\beq
Q \equiv c \cos q, \qquad P \equiv c \cos p 
\label{eq:defPQ}
\eeq
and
\beq
R \equiv (Q+P)/2, \qquad S \equiv t^{1/2}(Q-P)/2.
\label{eq:defRS}
\eeq
Inversely we have $P,Q=R\pm St^{-1/2}$, which
may be seen as the wavevector counterpart of
Eq.\,(\ref{eq:ijscaling}). 

\subsection{The poles of $g_{i,j;j'}(z;q,p)$}
\label{subsection:poles}

We first consider the $z$ integral in 
Eq.\,(\ref{eq:explGa}) and study the analytic structure of $g_{i,j;j'}(z;q,p)$
in the complex $z$ plane.
It may be shown that the various branch cuts that are present in the explicit expression (\ref{eq:explgM}) give no contribution after integration over $q$ and $p$. Indeed the only square roots come from the diagonalization of $\bF$ which is required to compute $(\bI \otimes \bI -\bF \otimes \bF)^{-1}$. The inverse of a general invertible matrix $\bN$ is given by $\bN^{-1} = (\det \bN)^{-1}\, (\cof \bN)^T$, where $(\cof \bN)$ denotes the matrix of cofactors. This shows that the coefficients of $(\bI \otimes \bI -\bF \otimes \bF)^{-1}$ are rational functions of the coefficients of $\bF$, which are themselves rational functions of $z$, involving no square roots.

Let $g_{i,j;j'}(z;q,p)$ have poles at $z_\sigma(R,S)$, where $\sigma$
is an index. Using the residue theorem we
may then cast the $z$ integral in (\ref{eq:explGa}) in the form
\beq
\frac{1}{2\pi\ci}
\oint_{\Gamma_0} \frac{\dd z}{z^t} \,g_{i,j;j'}(z;q,p) 
= \sum_{\alpha,\beta=\pm1} \sum_{\sigma} \alpha \beta\,
A_{\sigma}(R,\DQ) 
\ee^{\,t\cG_{\sigma,\alpha\beta}(R,\DQ;v,u)},
\label{eq:dvptzint}
\eeq
where we have written $\sin iq = (2\ci)^{-1}\sum_{\alpha = \pm 1} 
\alpha \ee^{\alpha \ci(vt+u\sqrt{t})q}$ 
and a similar expression for $\sin jp$,
the $A_{\sigma}$ are 
amplitudes whose dependence on $j'$ is not indicated explicitly, and 
the function in the exponential is defined by
\bea
\exp\Big( t\cG_{\sigma,\alpha\beta}(R,\DQ;v,u) \Big) &\equiv& 
z_{\sigma}^{-t} (\zeta_{\sigma}^2 a_{q\sigma} a_{p\sigma})^{vt}
\Big( \frac{a_{q\sigma}}{a_{p\sigma}} \Big)^{u\sqrt{t}} \nonumber \\
&& \times \,\ee^{\ci \alpha (vt+u\sqrt{t}) q} \,\ee^{\ci \beta (vt-u\sqrt{t}) p}, 
\label{eq:dcG}
\eea
in which $\zeta_\sigma$ and $a_{q\sigma}$ denote $\zeta$ and $a_q$ evaluated for $z=z_\sigma$, respectively, 
and $q$ and $p$ are to be expressed in terms of $R$ and $S$
through Eqs.\,(\ref{eq:defPQ}) and (\ref{eq:defRS}).

Once (\ref{eq:dcG}) is inserted in (\ref{eq:dvptzint})
which in turn is substituted in (\ref{eq:explGa}),
the $q$ and $p$ integrations in the latter equation have to be performed.
We will proceed on the hypothesis that these may be carried out by
means of a saddle point method, that is, that
for large $t$ 
these integrals will draw their main
contribution from narrow neighborhoods of saddle points 
$(R,\DQ) = (\Rs_{\sigma,\alpha\beta}(v,u),\DQs_{\sigma,\alpha\beta}(v,u))$ 
that are solutions of the coupled equations
\beq
\label{eq:saddle}
\frac{\partial\cG_{\sigma,\alpha\beta}(R,\DQ;v,u)}{\partial R} = 0, \qquad
\frac{\partial\cG_{\sigma,\alpha\beta}(R,\DQ;v,u)}{\partial \DQ} = 0.
\eeq
After the integrations on $q$ and $p$ are carried out,
we expect to find
that for $t\to\infty$ the Green function is dominated
by the term with the index $\sigma$ and the values of $\alpha$ and $\beta$ 
in (\ref{eq:dvptzint}) that have the largest saddle point value of 
$\Re\cG_{\sigma,\alpha\beta}$.
We will call this the `dominant saddle point' and refer to 
the pole that leads to it as the `dominant pole'. 
\vspace{2mm}

We now need to determine the poles $z_\sigma$ explicitly.
A high-order pole at $z = \rho^{-1}$
comes from the factor $\zeta^{i+j-j'}$. From equation (\ref{eq:explE}) we however see that the divergence at $z=\rho^{-1}$ does not come from the interaction between the two species $\pe$ and $\pn$. Rather, it is linked to the fact that mass would accumulate on a single site if the density of the traffic, that determines the probability to be blocked, was renormalized too heavily. This phenomenon is very generic and consequently cannot be at the origin of the pattern formation we seek to explain, thus discarding the pole at $z=\rho^{-1}$. The remaining poles
are located at the roots 
of 
\beq
\label{eq:polesLM}
1- \Bigg( \frac{\zeta^2 a_q a_p}{c^2} \Bigg)^2 = 0
\eeq 
or, equivalently, of
\beq
\label{eq:poleslm}
\zeta^2 a_q a_p = \epsilon c^2, \qquad  \epsilon = \pm 1.
\eeq
Let
\beq
\label{eq:defY}
Y \equiv {c^2}{\zeta}^{-1} = (\rho z)^{-1} - 1.
\eeq
We may deduce from
(\ref{eq:poleslm}) 
two polynomial equations in $Y$ by twice isolating the square roots 
in one of the members and squaring.
It then follows that the
$Y_\sigma \equiv c^2{\zeta_\sigma}^{-1}=(\rho z_\sigma)^{-1}-1$ are among
the roots of the two fourth-order polynomial equations 
\beq
\label{eq:polespoly}
Y^4 - 2 (1 + 2 \epsilon Q P ) Y^2 + 4 (Q^2+P^2) Y 
+ (1 - 4 \epsilon Q P) = 0,
\quad \epsilon = \pm 1.
\eeq
The analytical expressions of these roots
for general $Q$ and $P$ are of no practical use here. 
Instead, as anticipated by the scaling (\ref{eq:defPQ})-(\ref{eq:defRS}),
our analysis below will show that in the limit of large times
$t$ it suffices to know the solutions of (\ref{eq:polespoly}) in a
strip of width $\sim t^{-1/2}$ along
the diagonal $P=Q$, where the roots are easily found perturbatively.

\subsection{Selecting the dominant pole}
\label{subsection:dompole}

Finding out which 
one among the $z_{\sigma}$ is the dominant pole 
is not an easy task for general $i$ and $j$ but
can be done fairly easily 
in the special case where $i=j$ and hence, 
by Eq.\,(\ref{eq:ijscaling}), $u=0$. 
In that case $\DQs_{\sigma\alpha\beta}=0$ always solves the second one of the saddle point equations (\ref{eq:saddle}) by symmetry,
and we will suppose that this solution leads to the dominant saddle point.
Below we will find 
the corresponding $\Rs_{\sigma\alpha\beta}$
and see which set of indices $\alpha, \beta$, and $\sigma$
leads to the dominant saddle points.
We will then invoke continuity in $u$ to argue that
for $u\neq 0$ the same pole remains dominant 
and follows the path of the associated saddle points 
when they move off the $\DQ=0$ axis.
\vspace{2mm}

In the case $\DQ = 0$ we have $P=Q=R$ and the roots of Eqs.\,(\ref{eq:polesLM}) may be found explicitly. For fixed $\epsilon$, 
Eq.\,(\ref{eq:polespoly}) with $S=0$
has a double root
$Y_\epsilon=\epsilon$ and two further roots
$Y_{\epsiloneta} \equiv -\epsilon + 2 \eta \epsilon^{1/2} R$ 
where $\eta = \pm 1$ and the square root is defined everywhere with its
branch cut just below the negative real axis.
While Eq.\,(\ref{eq:polespoly}) is a necessary condition
that the roots of (\ref{eq:poleslm}) should satisfy, 
we still have to check if the roots found here actually do
solve Eq.\,(\ref{eq:poleslm}). This eliminates $Y_\epsilon$ as a
solution. Furthermore, each 
of the $Y_{\epsiloneta}$ solves Eq.\,(\ref{eq:poleslm}) with
$P=Q=R$ if and only if
certain conditions on $R$ are satisfied.
\vspace{2mm}

\begin{table}{}
\begin{center}
\begin{tabular}{|rr|r|c|c|c|c|}
\hline
$\epsilon$ & $\eta$ & $Y_{\epsiloneta}$\phantom{2R} & 
$\zeta_{\epsiloneta} = \frac{c^2}{Y_{\epsiloneta}} $  & 
$z_{\epsiloneta} = \frac{1}{\rho (1+Y_{\epsiloneta})} $ & 
Solves Eq.\,(\ref{eq:poleslm}) iff & Saddle points? \\ 
\hline
$1$ & $1$ & $-1+2 R$  & $- \frac{c^2}{1-2R}$ & $\frac{1}{2 \rho R}$ &
$\Im R > 0$& no\\ 
$1$ & $-1$ & $-1-2 R$  & $ -\frac{c^2}{1+2R}$ & $-\frac{1}{2 \rho R}$
& $\Im R < 0$ & no\\  
$-1$ & $1$ & $1+2\ci R$ & $\frac{c^2}{1+2\ci R}$ & $\frac{1}{2 \rho
  (1+\ci R)}$ & $\Im R > 1$ & no\\ 
$-1$ & $-1$ &  $1-2\ci R$ & $\frac{c^2}{1-2\ci R}$ & $\frac{1}{2 \rho
  (1 - \ci R)}$ & $\Im R > -1$ & $\Rs_\pm$\\ 
\hline
\end{tabular}
\end{center}
\caption{Values of $Y_{\epsiloneta}$, $\zeta_{\epsiloneta}$, 
  and $z_{\epsiloneta}$ for each solution of
  Eq.\,(\ref{eq:poleslm}) with $S=0$. Each expression is a solution only
  in a restricted domain of the complex $R$ plane indicated by the
  next to last column. The last column shows the possible saddle
  points of the function $\cG_{\epsiloneta,\alpha\beta}(R,0;v,0)$ in the
  domains of validity, that are expected to dominate the long time
  behaviour of the Green function.} 
\label{fig:tablepoles} 
\end{table}

We have thus found four solutions $Y_\sigma = Y_{\epsiloneta}$ (with
$\epsilon,\eta=\pm1$) to Eq.\,(\ref{eq:poleslm}) and will write the
corresponding values of $\zeta$ and $z$ as $\zeta_{\epsiloneta}$ and
$z_{\epsiloneta}$. All four have been listed in Table
\ref{fig:tablepoles}, together with the conditions on $R$.
When $u=0$, we ensure that the second one of the saddle point equations (\ref{eq:saddle})
is verified by setting $\DQs_{\epsiloneta,\alpha\beta}=0$,
and the first one of them becomes
$\partial\cG_{\epsilon\eta,\alpha\beta}(R,0;v,0)/\partial R = 0$ with
\bea\label{eq:gehab} 
\cG_{\epsilon\eta,\alpha\beta}(R,0;v,0) &=& 2v \log c + v \log \epsilon
+\ci (\alpha+\beta) v \arccos \bigg( \frac{R}{c} \bigg) - \log
z_{\epsiloneta} \nonumber\\
&=& (2-\alpha-\beta) v \log c + v \log \epsilon - \log(1+c^2) \\
&& + \log (1-\epsilon + 2 \eta \epsilon^{1/2} R) + (\alpha+\beta) v
\log(R + \ci (c^2-R^2)^{1/2}). \nonumber 
\eea
Examining the stationarity condition
shows that only for $\epsilon = \eta = -1$ and
$\alpha = \beta = -1$
do there exist saddle points in the complex $R$ plane
compatible with the conditions of Table \ref{fig:tablepoles}. 
Hence, for $u=0$  the dominant
pole $z_*(R,S)$ that leads to the final result is 
$z_*(R,0)=z_{-1,-1}(R,0)=1/[2\rho(1-\ci R)]$, 
given in Table \ref{fig:tablepoles}. 
Having singled out this pole we 
suppress the multiple indices $-1$ and write
$\cG_{-1,-1,-1,-1}(R,\DQ;v,u)=\cG(R,\DQ;v,u)$ and 
$A_{-1,-1}(R,\DQ)=A(R,\DQ)$.
The contribution of this pole to the $S$ integration
comes from the neigborhood of a saddle point
on the axis $S=\Ss_{-1,-1,-1,-1}=0$, for which
Eq.\,(\ref{eq:gehab}) can be made more explicit,
\bea
\cG(R,0;v,0) 
&=& 4v \log c + v \log \epsilon - \log(1+c^2) + \log 2\nonumber\\
&& + \log (1 - \ci R) -2v \log(R + \ci (c^2-R^2)^{1/2}).  
\label{eq:gehabbis} 
\eea
It has a pair 
of complex saddle points that we will denote by
$\Rs_\theta(v,0)$ with $\theta=\pm 1$. They obey
$(c^2-R^2)^{1/2} = 2 v (1- \ci R)$ and are therefore given by
\beq
\Rsth(v,0) = \Rth(v)\,, 
\qquad \theta = \pm1,
\label{eq:rsaddle}
\eeq
in which we introduce abbreviations that will serve again later on,
\beq
\label{eq:defRth}
\Rth(v) \equiv \frac{4 \ci v^2 + \theta \V}{1-4v^2},
\eeq
and
\bea
\label{eq:defU}
\V(v) &\equiv& (c^2 - 4v^2(1+c^2))^{1/2}\\[2mm]
   &=& \rho^{-1/2}(1-\rho-4v^2)^{1/2}.\nonumber
\eea
In the case $u=0$, the wavenumber integrations will
draw their dominant contributions from a narrow neighborhood
of one or both of 
the points $(R,\DQ)=(\Rth,0)$ with $\theta= \pm 1$,
depending on how the path of integration is routed.
We will consider this in the next sections, after extending the
discussion to the case of general $v$ and $u$.

\subsection{General expression for $\cG(R,\DQ;v,u)$}
\label{subsection:general}

We now consider the general case with $i$ and $j$ given by
Eq.\,(\ref{eq:ijscaling}), that is, $u \neq 0$, and we recall that $q$ and $p$ are linked to $R$ and $S$ via
Eqs.\,(\ref{eq:defPQ}) and (\ref{eq:defRS}).
We assume by continuity that the root 
$z_*(R,\DQ)$ identified above will 
continue to determine the final result for the Green function also
when $u\neq 0$.
Since each occurrence of $S$ is accompanied by a power $t^{-1/2}$
we may, for $t\to\infty$, expand $A(R,S)=A(R)+O(t^{-1/2})$, where
$A(R)\equiv A(R,0)$. 
The dominance of the pole at $z_*$ then allows us to simplify
Eq.\,(\ref{eq:dvptzint}) to
\beq
\frac{1}{2\pi\ci}
\oint_{\Gamma_0} \frac{\dd z}{z^t} \,g_{i,j;j'}(z;q,p) 
\simeq A(R)\exp\Big( \,t\cG(R,\DQ;v,u) \Big),
\qquad t\to\infty,
\label{eq:zint}
\eeq
which when substituted in (\ref{eq:explGa}) leads to
\beq
\Gee_{i,j;j'}(t) \simeq \int_{0}^{\pi}\!\dd q \int_{0}^{\pi}\!\dd p\, 
A(R) \exp\Big( \,t\cG(R,\DQ;v,u) \Big),
\qquad t\to\infty,
\label{eq:explGabis}
\eeq
where we have absorbed various factors in a redefinition
of the amplitude $A(R)$.

We first use Eq.\,(\ref{eq:polespoly}) to compute the expansion of
the pole $z_*(R,\DQ)$ perturbatively for small $\DQ$, 
knowing that each power of
$\DQ$ in this expansion is accompanied by a power of $t^{-1/2}$. 
The result is that 
\beq
\label{eq:zetam1m1}
\zeta_* = \frac{c^2}{1-2 \ci R 
           - \frac{\ci R \DQ^2}{(1-\ci R)^2t}} + O(t^{-2}), 
\eeq
\beq
\label{eq:zm1m1}
z_* = \frac{1}{2\rho\big( 1 - \ci R - \frac{\ci R \DQ^2}
             {2(1-\ci R)^2t} \big)} + O(t^{-2}). 
\eeq
From (\ref{eq:zetam1m1}) and (\ref{eq:defaq}) 
we also have the intermediate result
\beq
\label{eq:dvptam1m1}
\zeta_* a_{q*} = \ci c + \frac{c \DQ}{(1-\ci R) \sqrt{t}} + O(t^{-1}).
\eeq
in which $a_{q*}$ stands for $a_q$ evaluated at the pole.
Because of symmetry the expansion
of $\zeta_* a_{p*}$ can be obtained by replacing $\DQ$ with $-\DQ$. 
By inserting (\ref{eq:zetam1m1})-(\ref{eq:dvptam1m1}) in (\ref{eq:gehab})
for $\alpha=\beta=\epsilon=\eta=-1$ we obtain 
the large-$t$ expansion of $\cG$, 
which reads explicitly
\bea
\label{eq:exprginit}
\cG(R,\DQ;v,u) &=& \log 2 - \log (1+c^2) + 2 v \log c + \ci v \pi - \ci
v (q+p) \nonumber \\ 
&& + \log \bigg( 1 - \ci R - \frac{\ci R \DQ^2}{2 (1 - \ci R)^2t} \bigg)
+ \ci \frac{u}{\sqrt{t}} (p-q) \nonumber \\  
&& - \frac{2 \ci \DQ}{1-\ci R} \frac{u}{{t}} + O(t^{-2}) \nonumber \\ 
&= & \log 2 - \log (1+c^2) + 2 v \log c + \ci v \pi - 2 \ci v \arccos
\bigg( \frac{R}{c} \bigg) \nonumber \\ 
&& + \log(1-\ci R) + 2 \ci \frac{\DQ u}{{t}} 
\bigg( \frac{1}{\sqrt{c^2-R^2}} - \frac{1}{1-\ci R} \bigg) \nonumber \\  
&& -\ci \frac{R\DQ^2}{t} \bigg( \frac{1}{2(1-\ci R)^3} - \frac{
  v}{(c^2-R^2)^{3/2}} \bigg) + O(t^{-2}). 
\label{eq:exprgdvpt}
\eea 

The equations for the saddle points of the $R$ and $S$ integrations
are now coupled.
Solving them perturbatively in $t^{-1/2}$ we obtain
two pairs of saddle points $(\Rsth,\Ssth)$ given by 
\beq\label{eq:rsaddlegal}
\Rsth(v,u) = \Rth(v) + O(t^{-1})\,, 
\qquad \theta = \pm1, \\[2mm] 
\eeq
\beq
\label{eq:dqsaddlegal} 
\Ssth(v,u) = \frac{-4uv}{1+2v}
\frac{[1-\ci \Rth(v)]^2}{\Rth(v)} + O(t^{-1}), \qquad \theta=\pm 1,
\eeq
where $\Rth$ is defined in Eq.\,(\ref{eq:defRth}) and does not depend on $u$.

Having found the two pairs of saddle points $(\Rsth,\Ssth)$ indexed by $\theta=\pm1$ we will be able to carry out the integrations in the $R$ and $S$ planes. Performing a Taylor expansion of $\cG(R,\DQ;v,u)$ around these saddle points turns the integrals into Gaussian integrals in the $t \rightarrow \infty$ limit. We write $\cGsth(v,u) \equiv \cG(\Rsth,\Ssth;v,u)$. The integral along a contour that passes through $\Rsth$ and $\Ssth$ is then asymptotically evaluated to
\beq
\int\int_{(\Rsth,\Ssth)} \dd q\dd p\,A(R)\,\ee^{t\cG(R,\DQ;v,u)} = B_\theta\ee^{-\ci\phi_\theta}\,\frac{\exp [t\cGsth(v,u)]}{t}
\,[\,1\,+\,O(t^{-1/2})],
\label{eq:resultsaddle}
\eeq
where $B_\theta$ and $\phi_\theta$ are the amplitude and phase of the prefactor, and contain $A(\Rth)$, the second derivatives of $\cG$ with respect to $R$ and $S$, and the Jacobian $\partial (q,p)/\partial (R,t^{-1/2}\DQ)|_\theta$ that does not depend on time. We obtained expressions for these amplitudes but do not present them here. 

Our next task is to determine how the path of integration should be deformed in the $R$ and $\DQ$ planes. From the explicit expression (\ref{eq:exprgdvpt}) it is clear that minimizing $\cG(R,\DQ;v,u)$ over $\DQ$ gives a single saddle point for any value of $R$, so that finding the optimal path of integration in the $\DQ$ plane is straightforward once $R$ is fixed. In section~\ref{subsection:ondiag} we will identify the most convenient path in the $R$ plane depending on the value of $v$. The properties of the Green function will then be deduced form the functions $\cGsth(v,u)$ first on the diagonal and then in its vicinity.

\subsection{Green function on the diagonal: $i=j$}
\label{subsection:ondiag}

In this subsection we take $u=0$, so that $i = j = vt$ and $\Ssth = 0$.
By varying $v$ we therefore scan the Green function along the diagonal. 
By inspecting the variable $\V$ defined in (\ref{eq:defU}) 
we see that the velocity $v$ has a critical value
\beq
\label{eq:defvc}
\vc \equiv \frac{1}{2}\sqrt{c^2/(1+c^2)} = \frac{1}{2}\sqrt{1-\rho}
\eeq
below which $\V$ is real and above which it is pure imaginary.
We will dicuss these two cases separately.
\vspace{2mm}

{\it Case $v<\vc$.\,\,\,} This will appear to be the main regime.
For $v < \vc$ the saddle points $\Rs_\pm$ given by (\ref{eq:rsaddlegal})
are symmetric with respect to the imaginary axis. 
It directly follows that $\cGs_+$ and $\cGs_-$ are complex
conjugate.
For $u=0$ we have from (\ref{eq:exprgdvpt}) together with
(\ref{eq:rsaddlegal})-(\ref{eq:dqsaddlegal}),
\bea
\label{eq:gsaddle}
\cGs_\theta(v,0) &= &\log 2 - \log(1+c^2) + 4v \log c 
- (1-2 v) \log(1-2v) - \log(1+2v) \nonumber \\ 
&&+ \log(1-\ci \theta \V) - 2 v \log(2v-\ci \theta \V). 
\eea
The path of integration of the variable $R$, 
which runs from $-c$\, to $c$ 
along the real axis, will therefore be deformed
into the complex $R$ plane in such a way
that it passes through both saddle points $\Rs_\pm$. 
In the case $v < \vc$ there are therefore
two complex conjugate contributions of the form
(\ref{eq:resultsaddle}),
having $B_\theta=B$ and $\phi_\theta=-\theta\phi$. 
When substituted in (\ref{eq:explGabis}) these lead to
\beq
\Gee_{i,i;j'}(t) \simeq 2B\,\ee^{t\Re\cGs_+(v,0)}
\cos(t\Im\cG_+(v,0)-\phi),  \qquad i=vt, \quad v<\vc\,, \quad t\to\infty.
\label{eq:greenfinal}
\eeq
The real and imaginary parts
of $\cGs_\pm$ are, from (\ref{eq:gsaddle}), 
\bea
\label{eq:regvinf}
\Re\cGs_\pm(v,0) &=& \log 2 - \frac{1}{2} \log(1+c^2) - \frac{1}{2} (1-
2v) \log(1-2 v) \nonumber \\  
&&- \frac{1}{2} (1+2v)\log(1+2v) + 2 v \log c, \qquad v<\vc\,,\\[2mm]
\Im\cGs_\pm(v,0) &=& \mp \big[ \arctan \V - 2 v \arctan (\V/2v) \big],
\qquad v<\vc\,.
\label{eq:imgvinf}
\eea
Upon casting the argument of the cosine in 
Eq.\,(\ref{eq:greenfinal}) in the form
$t\Im\cG_\pm(v,0) = \mp (\omega t - ik)$
we find the expressions
\beq
\omega(v) = \arctan \V, \qquad k(v) = 2\arctan(\V/2v), 
\label{eq:omegak}
\eeq
for the angular frequency $\omega$
and the wavenumber $\gk=2^{-1/2}k$\footnote{Distances are systematically reduced by a factor $\sqrt{2}$ when one deduces the properties of the wave packet in the diagonal direction from those along the $i$ or $j$ axis. In the following we overline the quantities in which this operation has been performed.
}, respectively;
they are
valid along the diagonal $i=j=vt$ for
variations $\Delta i\ll i$. 
This immediately yields
the wavelength $\glambda(v)$ of the oscillations as a function of the
velocity $v$ and the density $\rho$,
\beq
\label{eq:wavelength}
\glambda(v) = \frac{2 \pi}{\gk(v)} = \frac{\sqrt{2} \pi}{\arctan(\V/2v)},
\eeq
in which $\V$ is given by Eq.\,(\ref{eq:defU}). 

Eq.\,(\ref{eq:greenfinal}) shows that the Green function oscillates,
which constitutes the proof of the instability that we were looking
for. We will therefore sometimes refer to this Green function as a
`wave packet.' 

The complexity of the calculations presented here
begs for independent confirmation.
To that end we have 
applied an instantaneous perturbation at $t=0$
to a single boundary site, usually $(1,0)$ or $(0,1)$, 
and iterated the linearized equations (\ref{eq:l})-(\ref{eq:ic}) numerically
in time.
This leads to numerically exact values for the Green functions
that may be compared to the analytic results.
The numerical calculations were 
carried out on a lattice of linear size $M=800$ 
and the number of iterations in time varied between $400$ and $1200$.

\begin{figure}
\begin{center}
\scalebox{.50}
{\includegraphics{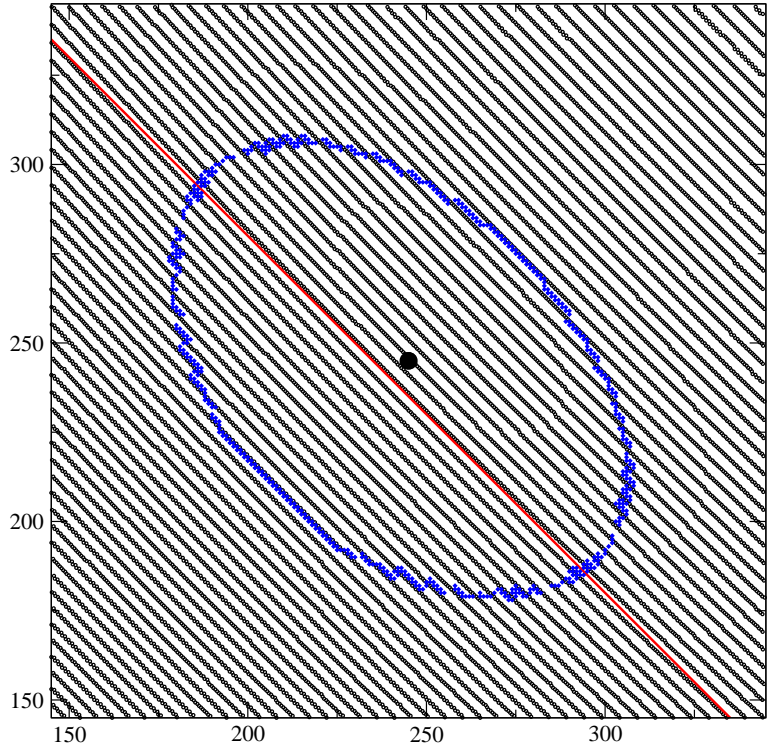}} 
\end{center}
\caption{\small Crests of the Green function $\Gee_{i,j;1}(t)$ for $\rho=0.3$ and $t=1200$. The local wavelength $\glambda(v)$ increases between the lower left and the upper right, as predicted by (\ref{eq:wavelength}). The crests are slightly concave (see section \ref{subsection:offdiag}), as can be seen by comparison with the straight red line $i+j=2 \vg t$. The amplitude is maximal at the black dot and one tenth of the peak amplitude on the blue curve.}  
\label{fig:redblack}
\end{figure}

In figure \ref{fig:redblack} we show the numerically determined crests
of one of the four Green functions in a square subregion of the lattice.
The wave packet amplitude (not shown) has its peak at $i=j\approx 245$.
On each line parallel to the
$(1,1)$ direction the values of the Green function have been
interpolated to determine the positions of the local maxima.
The stripe formation instability clearly appears.

\begin{figure}
\begin{center}
\scalebox{.50}
{\includegraphics{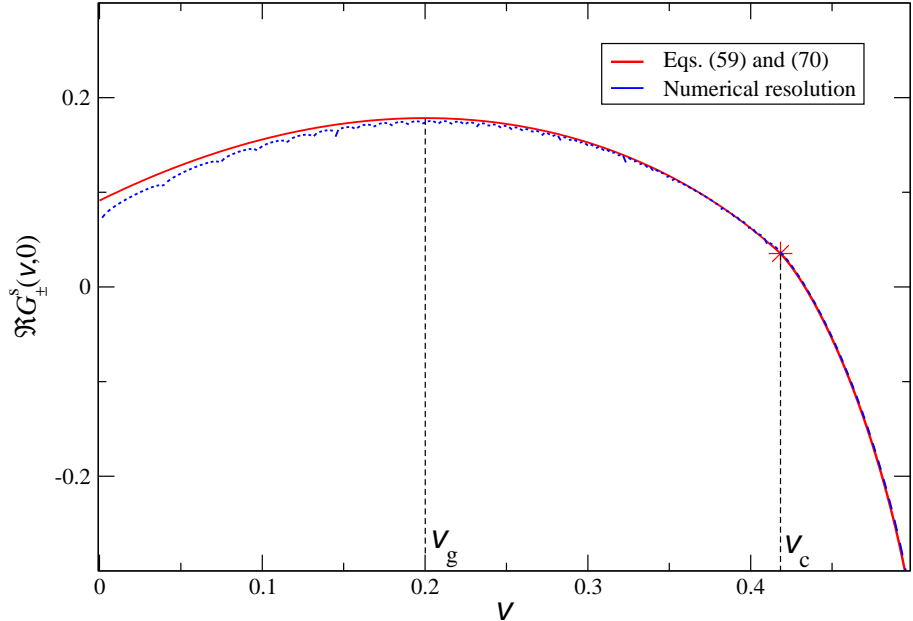}} 
\end{center}
\caption{\small The logarithm $\Re\cGs_\pm(v,0)$ of the envelope
  of the wave packet along the diagonal
  $i=j$ for $t=400$ and $\rho = 0.3$. Shown are
  both the analytical prediction (red) from Eqs.\,(\ref{eq:regvinf}) and
  (\ref{eq:gvsup}) and the numerical solution (blue) of the linearized
  evolution equations (\ref{eq:l})-(\ref{eq:ic}). The red star emphasizes a discontinuity in the slope.}  
\label{fig:envelope}
\end{figure}

The envelope of (\ref{eq:greenfinal}), determined by (\ref{eq:regvinf}),
peaks at a value $i=j=\vg t$ where $\vg$ is the solution
of $\partial \cGs_\pm/\partial v|_{\vg} =0$.
The value of $\vg$ is interpreted as the projection of the group velocity along the direction $i$ (or $j$), which gives for the true group velocity of the packet $\gvg = \sqrt{2} \vg$.
We find
\beq
\vg =\frac{c^2-1}{2(c^2+1)} = \frac{1}{2} - \rho.
\label{eq:rvg}
\eeq
This shows that our description makes sense, at best, in the density interval $0 < \rho < 1/2$.
Substitution of (\ref{eq:rvg}) in (\ref{eq:regvinf}) yields 
a remarkably simple expression for the maximal growth rate 
$\Re\cGs_\pm(\vg,0)$,
\bea
\label{eq:regvinfpeak}
\exp \Re\cGs_\pm(\vg,0)&=& (1+c^{-2})^{-1/2} \nonumber\\
&=& (1-\rho)^{-1/2},
\eea
so that $|\Gee_{i,i;j'}(t)| \sim (1-\rho)^{-t/2}$ for $i=\vg t$ and 
$t \rightarrow \infty$. 
This growth rate is identical to the one associated with
periodic boundary conditions (Ref.\,\cite{cividini_h_a2013}, subsection 4.2).

The amplitude of the Green function is shown in figure \ref{fig:envelope} as a function of
$v$, together with its numerical determination.  The prefactor has been adjusted to obtain the best agreement between both curves. The maximum occurs at $v=\vg=0.2$ [Eq.\,(\ref{eq:rvg})] and there is a discontinuity in the slope at $v = \vc \simeq 0.41833$ [Eq.\,(\ref{eq:defvc})], which is well brought out numerically.

At the maximum of the peak we have $v=\vg$. Using (\ref{eq:defU}) and
(\ref{eq:rvg})
to express $\glambda_{\rm 0}=\glambda(\vg)$ as a function only of 
$\rho$ we thus find from (\ref{eq:wavelength})
that this wavelength of maximal instability has the expression
\bea
\label{eq:wavelengthpeak}
\glambda_{\rm 0} 
&=& \sqrt{2} \pi / \arctan\Bigg( \frac{\sqrt{3-4 \rho}}{1-2 \rho} \Bigg)
\nonumber\\[2mm]
&=& \sqrt{2} \pi / \arccos\Bigg( \frac{1-2\rho}{2(1-\rho)} \Bigg).
\eea
Remarkably, this expression for the most unstable wavelength is
is identical to the one found in Ref.\,\cite{cividini_h_a2013} 
for the much simpler case of periodic boundary conditions.
This therefore points towards a robust property of the mean field
equations. 
Typically, $\glambda_{\rm 0}$ is of the order of three to
four lattice spacings.
Let $\gvph=\omega(\vg)/\gk(\vg)$
denote the phase velocity of the oscillations inside the
wave packet. From Eq.\,(\ref{eq:omegak}) we have
\beq
\label{eq:xvph}
\gvph = \frac{\arctan\sqrt{3-4\rho}}{\sqrt{2}\arccos\bigg(\frac{1-2\rho}{2(1-\rho)}\bigg)}\,.
\eeq

\begin{figure}
\begin{center}
\scalebox{.50}
{\includegraphics{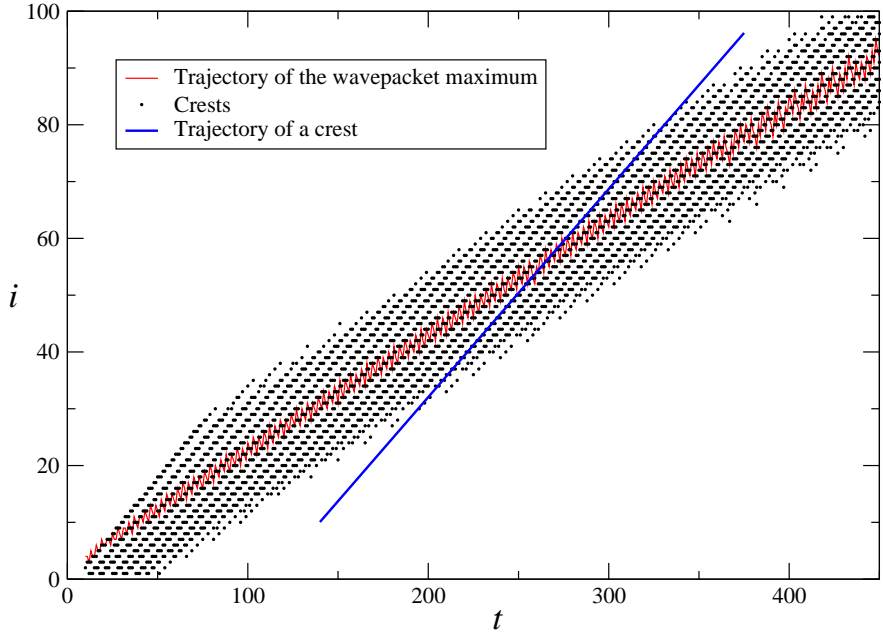}} 
\end{center}
\caption{\small Numerical determination of the crests of the Green function $\Gee$ for $\rho=0.30$ shown in the time-space plane.
  Positions are shown only of those crests that are within a distance
  of ten lattice units from the wave packet maximum.}  
\label{fig:crests}
\end{figure}

\begin{figure}
\begin{center}
\scalebox{.45}
{\includegraphics{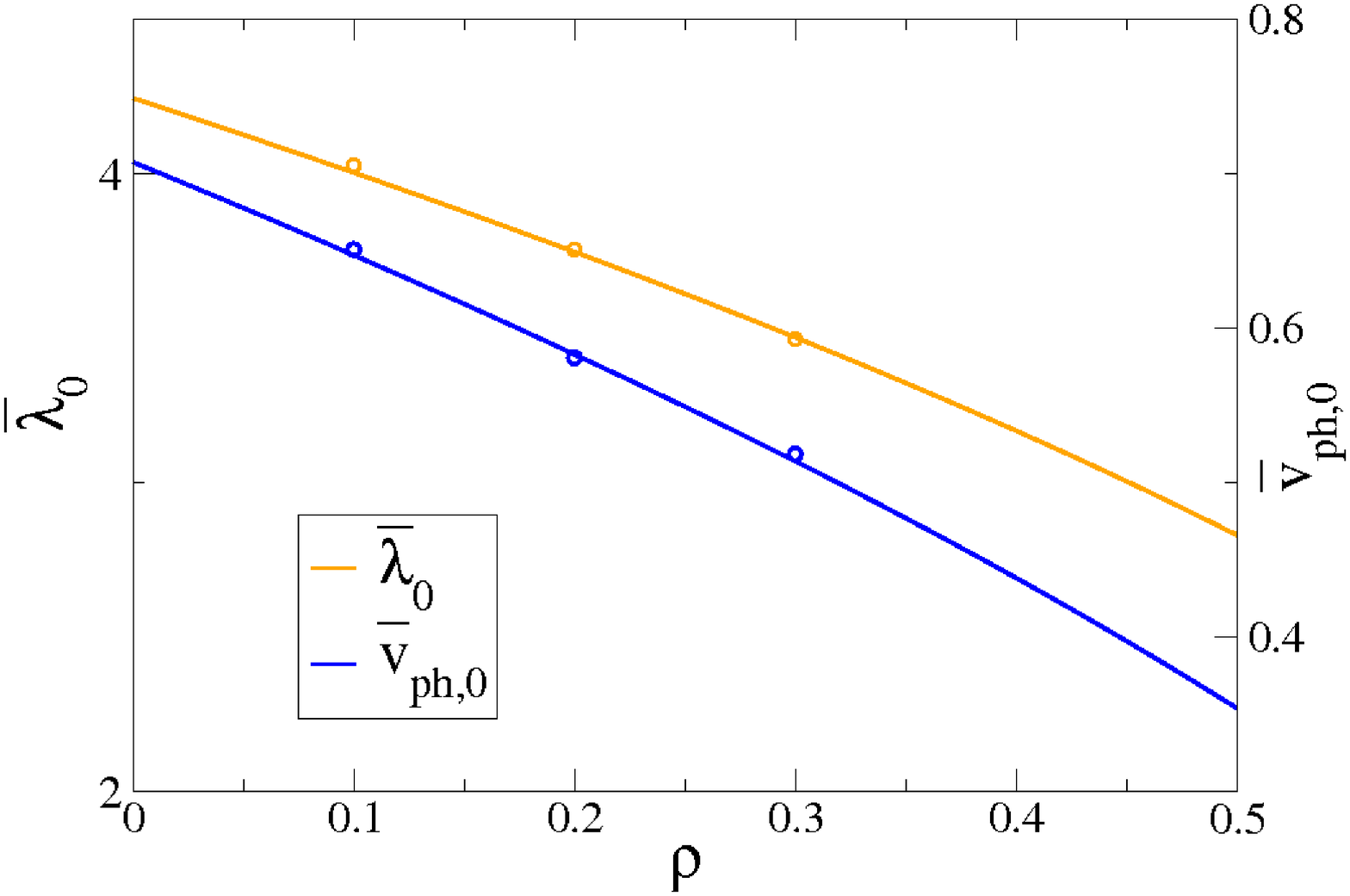}} 
\end{center}
\caption{\small Wavelength $\glambda_{\rm 0}$ and phase velocity $\gvph$ at the maximum of the peak as a function of $\rho$. The lines are the analytic predictions and the dots from numerical determination. The uncertainty comes from the estimation of $\glambda_{\rm 0}$ and $\gvph$ from the numerical data, and in both cases the error bars are smaller than the symbol size.}  
\label{fig:imcg}
\end{figure}

Figure \ref{fig:crests} is based on a numerical determination of the
Green function for  $\rho=0.30$ along the diagonal $i=j$.
The fluctuating (red) solid line represents the trajectory 
of the maximum of the wave packet in the $ti$ plane.
Its fluctuations  are due to the incommensurability between the
wavelength of the oscillations and the lattice spacing. Its average
slope is the projected group velocity $\vg$. The black dots are the positions
of the crests, limited to those within a distance of ten lattice units
from the maximum. In this figure a crest is a lattice site $(i,j)$ such that $\drhoe_{i-1,j-1} < \drhoe_{i,j} > \drhoe_{i+1,j+1}$. The straight (blue) solid line is the trajectory 
of a given crest in the $ti$ plane. Its slope is the projection of the phase velocity. A few values of $\glambda_{\rm 0}$ and $\gvph$ determined from plots similar to figure \ref{fig:crests} are shown in figure \ref{fig:imcg}.

Finally, the width of the peak along the diagonal direction can be calculated as well. Defining
\beq
\label{eq:sigmaparpeak}
\sigma_{\parallel, {\rm 0}}^2 \equiv \frac{1-4 \vg^2}{4} = \rho (1-\rho),
\eeq
we have
\bea
\label{eq:defsigmapar}
t \Re\cGs_\pm(v,0) &=& t \Re\cGs_\pm(\vg,0) - t \frac{(v-\vg)^2}{2 \sigma_{\parallel, {\rm 0}}^2} + t O((v-\vg)^3)\nonumber \\
&=& t \Re\cGs_\pm(\vg,0) - \frac{((i+j-2 \vg t)/2)^2}{2 \sigma_{\parallel, {\rm 0}}^2 t} + O(t^{-2}),
\eea
which shows that the perturbation spreads diffusively. Eq.\;(\ref{eq:sigmaparpeak}) is the standard expression for the variance of a Bernoulli distribution of parameter $\rho$. It can be seen in the evolution equations (\ref{eq:l}) that a perturbation of the density field will increase its value of $i+j$ with probability $1-\rho$ and decrease this value with probability $\rho$ at each time step, which explains the expression (\ref{eq:sigmaparpeak}). This interpretation also explains why the expression (\ref{eq:rvg}) for $\vg$ becomes negative when $\rho > 1/2$. We have verified expression (\ref{eq:sigmaparpeak}) numerically.
\vspace{2mm}

{\it Case $v>\vc$.\,\,\,} 
When $v > \vc$ the quantity $\V$ becomes pure imaginary and we define
\beq
W \equiv -\ci V = \sqrt{4v^2(1+c^2)-c^2}, \qquad v>\vc\,,
\label{eq:xU}
\eeq
which is real positive, so that both saddle points $\Rsth$ are now pure imaginary.
We direct the path of integration
only through $\Rs_+$, since at this point 
the direction of negative curvature is parallel to the real axis whereas in $\Rs_-$ the two directions are perpendicular. 

Equations (\ref{eq:exprgdvpt}) together with
(\ref{eq:rsaddlegal})-(\ref{eq:dqsaddlegal})
lead to the counterpart of (\ref{eq:gsaddle}),
\bea
\label{eq:gvsup}
\cGs_+(v,0) &=& \log 2 -\log (1+c^2) + 4 v \log c - (1-2v) \log(1-2v) -
\log(1+2v) \nonumber\\ 
&&+ \log(1+W) - 2v \log(2v+W),  \qquad v>\vc,
\eea
The $\cGs_\pm$ are 
real and the Green function depends exponentially on $t$.
\beq
\Gee_{i,i;j'}(t) \simeq B_+\,\ee^{t\cGs_+(v,0)},
\qquad i=vt, \quad v>\vc\,, \quad t\to\infty.
\label{eq:gsaddlebis}
\eeq 

Eqs.\,(\ref{eq:gvsup}) and (\ref{eq:gsaddlebis}) together give the
envelope of the Green function. This second part of the envelope
has also been calculated
numerically and is also shown in figure \ref{fig:envelope}.
We note that it concerns the front of the wave, where the amplitude is
still extremely small compared to its peak value.  
Numerical and analytic work are in excellent agreement.


\subsection{Green function off the diagonal: $i \neq j$}
\label{subsection:offdiag}

In this last subsection we study the
Green function for the case $u \neq 0$.
At fixed $v$, varying $u$ corresponds to scanning the Green
function along an `antidiagonal'
of constant $i+j$.
The most interesting case is when $i+j=2\vg t$,
which is the antidiagonal passing through the peak of the wave packet.

For $u\neq 0$ and $v$ arbitrary the function to study
is then the full expression~(\ref{eq:exprgdvpt}). 
Substitution of Eqs.\,(\ref{eq:rsaddlegal}) and (\ref{eq:dqsaddlegal}) in
Eq.\,(\ref{eq:exprgdvpt}) gives again $\cGsth(v,0)$ of
Eq.\,(\ref{eq:gsaddle}) but
augmented with terms of order $u^2/t$. Explicitly,
\bea
\label{eq:gsaddleu}
\cGsth(v,u) &=& \cGsth(v,0) +
\frac{2(1-2v)}{1+2v} \frac{1-\ci \Rth}{\ci \Rth}
\frac{u^2}{t} + O(t^{-2}) \nonumber\\ 
&=& \cGsth(v,0) - \frac{1}{2}\left( \frac{1}{\sigma_{\perp}^2}+\ci \theta \phi^\pp \right)
\frac{u^2}{t} + O(t^{-2}).
\eea  
The real term gives the transverse width of the wave packet. Using the explicit expression (\ref{eq:defRth})  
for $\Rth$ we find
\beq
\sigma_{\perp}^2 = \frac{c^2-4v^2}{4c^2}\frac{1+2v}{1-2v}.
\label{eq:xsigma2}
\eeq
The imaginary term is in fact the second derivative of the phase of the cosine in equation (\ref{eq:greenfinal}) with respect to $u$ and reads
\beq
\phi^{\pp} = \frac{4V(1-2v)}{(1+2v)(c^2-4v^2)}\,,
\label{eq:xphipp}
\eeq
in which $V$ is given by (\ref{eq:defU}).
By combining again the results from the two saddle points 
$\theta=\pm 1$
we find that, to second order in $u$ around the diagonal,
the wave packet is given by a generalization of (\ref{eq:greenfinal}),
\bea
\label{eq:greenfinalu}
\Gee_{i,j;j'}(t) &\simeq& 2B\,\ee^{t\Re\cGs_+(v,0)}\ee^{-u^2/(2\sigma_{\perp}^2)}
\cos \bigg(\omega t - \frac{1}{2}(i+j)k + \phi + \frac{\phi^{\pp}}{2}u^2\bigg),  
\nonumber\\[2mm]
&& i,j=vt\pm ut^{1/2}, \quad v<\vc\,, \quad t\to\infty,
\eea
where $\Re\cGs_+(v,0)$, $\omega(v)$, and 
$k(v)$ are given by expressions
(\ref{eq:regvinf})and (\ref{eq:omegak}) of
subsection~\ref{subsection:ondiag}. 
By substituting in (\ref{eq:xsigma2}) and
(\ref{eq:xphipp}) the expressions for
$\vg$ and $c$ in terms of $\rho$ one finds, in obvious notation, 
\beq
\label{eq:sigmaperppeak}
\sigma_{\perp,{\rm 0}}^2 = \frac{1-2\rho+4\rho^2-4\rho^3}{4\rho} = \rho (1-\rho) + \frac{1-2\rho}{4\rho}\,. 
\eeq
and
\beq
\label{eq:phaseshiftpeak}
\phi^{\pp}_{\rm 0} = \frac{4 \rho^2 \sqrt{3-4\rho}}{(1-\rho) (1-2\rho +4\rho^2-4\rho^3)}\,. 
\eeq
From equations (\ref{eq:sigmaparpeak}) and (\ref{eq:sigmaperppeak}) we notice that the variance in the antidiagonal direction $\sigma_{\perp,{\rm 0}}$ is always larger than its diagonal counterpart, in accordance with the scheme drawn in figure \ref{fig:wavepacket}. According to Eq.\,(\ref{eq:greenfinalu}) the Green function should
vanish every time the cosine in that equation has a zero.

\begin{figure}
\begin{center}
\scalebox{.50}
{\includegraphics{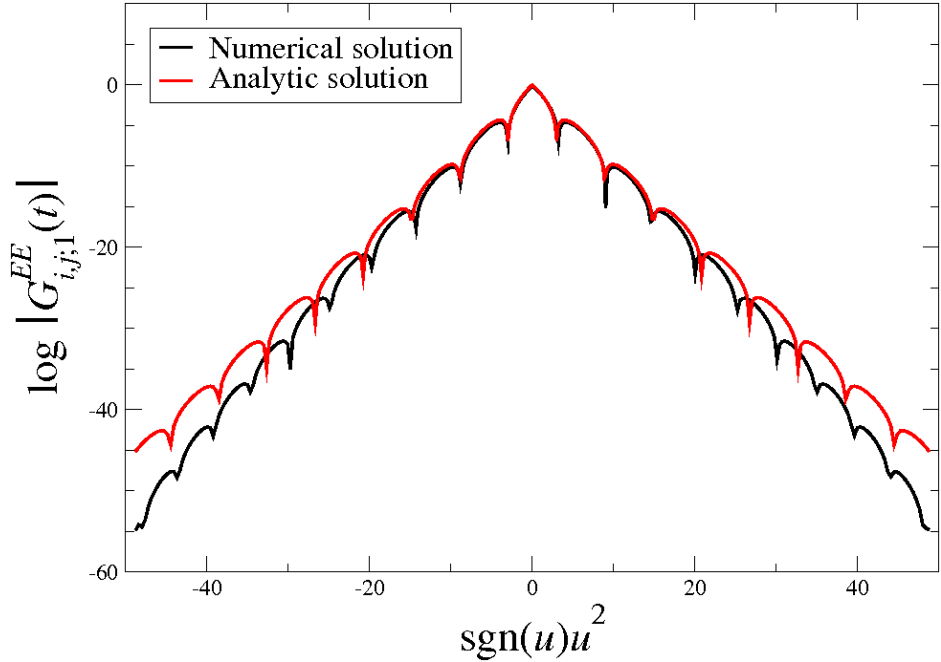}} 
\end{center}
\caption{\small Logarithm of the Green function
$\Gee_{i,j;1}(t)$  for $i,j=\vg t \pm  u \sqrt{t}$,
evaluated for  $\rho=0.30$ at time $t=1200$.
The black curve is the numerical solution and the red one is the analytic expression (\ref{eq:greenfinalu}). Both curves have been shifted so that they have unit amplitude in $u=0$, and the value of $\phi$ in the analytic expression has been fitted.
}  
\label{fig:antiprofile}
\end{figure}

In figure \ref{fig:antiprofile} we show the Green function $\Gee_{i,j;1}$ along the antidiagonal $i+j=2 \vg t$, that passes through the peak of the wave packet, as a function of $u = (i-j)/(2\sqrt{t})$. The downward dips are the divergences of $\log |\Gee_{i,j;1}|$ that occur when the cosine in (\ref{eq:greenfinalu}) vanishes. There is excellent agreement for smal $u$ values; since the analytic expression is based on a small-$u$ expansion, it is normal that for larger $u$ deviations appear.

Whereas the term $(i+j)k/{2}$ in the argument of the cosine in
(\ref{eq:greenfinalu}) suggests the propagation of plane waves in the
$(1,1)$ direction, the term $\frac{\phi^{\pp}}{2} u^2$ induces a very slight curvature, which is nevertheless clearly visible in figure \ref{fig:redblack}.

This effect is theoretically very interesting and we are not aware of any intuitive explanation. Since it appears 
in the wings of the wavepacket where the amplitude is very small,
it will be washed out when the Green function
is convoluted with the space and time dependent noise at the entrance 
boundaries. It might however be observable under idealized circumstances with a pure 'instanton' perturbation of a homogeneous flow. We note that this curvature effect is not related to a different curvature
phenomenon, termed the {\it chevron effect,} that we 
discovered and described in earlier work 
\cite{cividini_a_h2013,cividini_h_a2013} and that appears once 
a stationary state has set in under the influence of the nonlinear
terms in the equations. The chevron effect is essentially nonlinear; no
evidence of it was found in the initial linear regime studied in this
work.

This completes the asymptotic analysis of the Green function
$\Gee$. The same asymptotic arguments can be applied to
$\Gen$, which differs from $\Gee$ only by a factor
$\big(\frac{\zeta a_p}{c}\big)^2$ in the integrand
and a relabeling of the indices.
In particular the
shape of the peak and the wavelength of the oscillations are the same
as those found for $\Gee$. The two other Green functions $\Gne$ and
$\Gnn$ are obtained by exchanging $(\pe,i)$ with $(\pn,j)$. 

\section{Summary and conclusion}
\label{section:ccl}

We have studied the stripe formation instability known to occur
in the crossing area of two perpendicular
traffic flows (`eastward' and `northward')
through streets of suffcient width.
The phenomenon is common to a wide class of models.
In the present work the two streets were modeled as strips of a square
lattice of width $M$.
We have started from the deterministic nonlinear
mean-field flow equations (\ref{eq:nl})
whose unknowns are the space and time dependent
densities $\rhoe_{i,j}$ and $\rhon_{i,j}$ of the eastbound and
northbound traffic, respectively. 
These nonlinear equations cannot be solved analytically.
In earlier work \cite{cividini_h_a2013} we therefore
performed a Monte Carlo
study of the  stationary states of these equations, which are unstable to the appearance of a fully developed striped pattern.
Here our purpose has been to study the initial {\it linear\,}
growth of this
instability and to show that it may be triggered by random open boundary
conditions (OBC), representing randomly incoming traffic at the west and south entrance boundaries of the crossing area.
The same instability was also studied analytically
(\cite{cividini_h_a2013}, section 4) in the more artificial geometry of periodic boundary conditions (PBC) and subject to a random
initial condition.
For the linear problem (\ref{eq:l}) at hand all information is contained 
in the Green functions. 
Their expressions can
be calculated exactly {\it via\,} 
diagonalization of the time evolution matrix
and are given by Eqs.\,(\ref{eq:explGaM})-(\ref{eq:explGbM}) as a double sum of an integral.
We have evaluated these asymptotically in the limit of large times
and shown that the Green function represents a wave packet
of growing amplitude that propagates 
along the $(1,1)$ direction.

For the traveling wave packet generated by an instantaneous
point-like disturbance on one of the boundaries, we found explicit expressions for the wavelength $\glambda_{\rm 0}$
of maximum instability as a function of the average traffic density $\rho$,
for the growth rate of the instability, and for the group and phase
velocities, $\gvg$ and $\gvph$, respectively.
We found full agreement between our analytic results and  a numerical
determination of the Green function.

We concluded that random entrance boundary conditions 
(OBC) generate a wave pattern similar to the one found under the much
simpler PBC.
This result is interesting and important for the analysis of
similar models. It shows that the simplified version of a model with
PBC and random initial conditions may quite well replace the full model as far as the features
studied here are concerned. 
Nevertheless, the calculation of the Green function of the full model with OBC,
as carried out in this work, gives a much deeper insight into the 
interaction between the two crossing flows,
the selection mechanism of the dominant mode of propagation,
and the different time scales involved.
It has brought to light an interesting
curvature effect of the wavefront in the wings of the wave packet, created
by an instanton perturbation.
Furthermore, the knowledge about the Green function that we have
acquired here
may be applied directly, by means of a simple convolution,
to more complicated cases with arbitrary boundary conditions
in which the entrance noise may or may not have correlations.
It is very likely, moreover, that our
method can be extended to various other situations that may arise.

\section*{Acknowledgment}

The authors thank Professor Martin R. Evans for a discussion.

\end{document}